\documentclass[aps,prd,twocolumn,notitlepage,nofootinbib,floatfix,superscriptaddress]{revtex4-1}
\usepackage{amsmath, amsthm, amssymb, amsfonts, amsbsy,mathrsfs}

\usepackage{xcolor}
\usepackage{calligra,bm}
\usepackage{stmaryrd}
\usepackage{graphicx}
\usepackage[T1]{fontenc}
\usepackage{soul}

\graphicspath{{./images/}}

\usepackage[hidelinks,bookmarks=true]{hyperref}
\hypersetup{pdfstartview=FitH,pdfhighlight=/O,colorlinks=false}

\usepackage[normalem]{ulem}
\def\ipartial{{\mathchar'26\mkern-12mu \partial}}

\begin{document}

%
%

\title{Almost-Killing equation: Stability, hyperbolicity, and black hole Gauss law}

  \author{Justin C. Feng}
	   \affiliation{CENTRA, Departamento de F{\'i}sica, Instituto Superior T{\'e}cnico - IST, Universidade de Lisboa - UL, Avenida Rovisco Pais 1, 1049 Lisboa, Portugal}
  \author{Edgar Gasper\'in}
     \affiliation{CENTRA, Departamento de F{\'i}sica, Instituto Superior T{\'e}cnico - IST, Universidade de Lisboa - UL, Avenida Rovisco Pais 1, 1049 Lisboa, Portugal}
  \author{Jarrod L. Williams}
	   \affiliation{School of Mathematical Sciences, Queen Mary, University of London, Mile End Road, London E1 4NS, United Kingdom.}

%
%
\begin{abstract}
We examine the Hamiltonian formulation and hyperbolicity of the almost-Killing equation
(AKE). We find that for all but one parameter
choice, the Hamiltonian is unbounded and in some cases, the AKE has
ghost degrees of freedom. We also show the AKE is only strongly
hyperbolic for one parameter choice, which corresponds to a case in
which the AKE has ghosts. Fortunately, one finds that the AKE reduces
to the homogeneous Maxwell equation in a vacuum, so that with the
addition of the divergence-free constraint (a ``Lorenz gauge'') one
can still obtain a well-posed problem that is stable in the sense that
the corresponding Hamiltonian is positive definite. An analysis of the
resulting Komar currents reveals an exact Gauss law for a system of
black holes in vacuum spacetimes, and suggests a possible measure of
matter content in asymptotically flat spacetimes.
\end{abstract}


\maketitle

%
%

%
%
\section{Introduction}
The construction of approximate Killing vectors is motivated by the prospect of constructing conserved quantities in spacetimes which lack symmetry. One such construction was first proposed by Komar in \cite{Komar1959,Komar1962}, where it was observed that given any vector $u^\mu$, the so-called Komar current (sometimes referred to as the Noether current in the literature---see for instance, \cite{Wald94,*IyeWal94,ChaPad15}),
\begin{equation} \label{AKEGL-KomarCurrent}
	J^\mu[u] := 2 \, \nabla_\nu \left( \nabla^{[\mu} u^{\nu]} \right) ,
\end{equation}
\noindent satisfies the divergence-free condition $\nabla \cdot J [u]=0$. We stress here that the divergence-free property of $J^\mu[u]$ is independent of the choice of the vector $u^\mu$. The divergence-free property of $J^\mu[u]$ permits the construction of conserved quantities which may be rewritten as surface integrals---in particular, the following integral over the spacelike hypersurface $\Sigma$ satisfies:
\begin{equation} \label{AKEGL-KomarSurfInt}
	Q=\int_{\Sigma} J^\mu[u] \, d\Sigma_\mu = 2 \int_{\partial \Sigma} \nabla^{[\mu} u^{\nu]}\, dS_{\mu \nu}.
\end{equation}
\noindent In \cite{Komar1962}, Komar studied currents and conserved quantities in asymptotically flat spacetimes considering vectors $\xi^\mu$ which asymptotically satisfy the Killing condition $\nabla_{(\mu}\xi_{\nu)} = 0$ as one approaches spatial infinity, termed ``semi-Killing'' vectors. Since then, there have been several attempts to construct approximate notions of Killing vectors, for instance the Eigenvector approach of Matzner \cite{Matzner1968,*Beetle2008,*BeetleWilder2014}, the symmetry-seeking coordinates of Garfinkle and Gundlach \cite{GarfinkleGundlach1999}, the affine collineation approach of Harte \cite{Harte2008}, and the almost-Killing equation (AKE) \cite{Taubes1978,Bonaetal2005}, the properties of which will form the main focus of this article. The study of Komar currents constructed from solutions of the AKE was examined in \cite{Ruizetal2014} and \cite{FengCurrents2018}, and we explore further the properties of Komar currents constructed from solutions of the AKE.

In particular, we show that for certain parameter choices or for divergence-free solutions, one can formulate an exact Gauss law in asymptotically flat spacetimes, provided that the Ricci tensor has compact support on spatial hypersurfaces. Furthermore, we show that this Gauss law applies to a system of black holes in vacuum spacetimes, and that one can construct an invariant quantity from the sum of surface integrals over black hole horizons. Of course, the reader may recall that a Gauss law was proposed in \cite{Komar1962} for semi-Killing vectors in asymptotically flat spacetimes, but one might expect that in general, a Gauss law constructed from semi-Killing vectors will only be approximate. What is new here is the finding that divergence-free solutions of the AKE (assuming they exist) in asymptotically flat spacetimes can yield an \textit{exact} Gauss law for black holes in vacuum and matter confined to finite regions in space. In principle, this can be used to obtain a generalization of the Komar mass (and in fact reduces to the Komar mass for Killing and semi-Killing vectors), but the resulting conserved charge is insensitive to gravitational radiation in vacuum spacetimes. For this reason, the generalization of Komar mass obtained from solutions of the AKE cannot provide a satisfactory notion of gravitational energy. However, we argue that it can still provide a useful measure of matter content in certain cases.

One might wonder whether solutions of the AKE can be used to construct long-lived semi-Killing vectors in asymptotically flat spacetimes. Motivated by this prospect, we examine the Hamiltonian formulation and hyperbolicity of the AKE, with the goal of identifying the cases where the AKE is bounded below and admits a well-posed Cauchy problem. A Hamiltonian analysis reveals a difficulty; the Hamiltonian is generally unbounded below and the AKE can have ghost degrees of freedom for certain parameter choices. This issue is problematic because an unbounded Hamiltonian generally signals the presence of runaway instabilities, which can potentially drive solutions far from the Killing condition. We show that there is one parameter choice for which the AKE is strongly hyperbolic; the requirement of strong hyperbolicity fixes the free parameter of the AKE \cite{Bonaetal2005,Taubes1978}, and also excludes the parameter choice for the ``conformal almost-Killing Equation'' introduced in \cite{Taubes1978}. Unfortunately, the parameter choice which makes the AKE strongly hyperbolic also introduces ghosts into the system, which at first sight seems to indicate that the AKE cannot simultaneously have a bounded Hamiltonian and admit a well-posed Cauchy problem. However, there is one instance in which this difficulty can be overcome. It can be shown that for a particular choice of initial data in vacuum \cite{Bonaetal2005}, the solutions of the strongly hyperbolic instance of the AKE can be identified with those of the homogeneous Maxwell equation for the four potential, supplemented with a constraint equation that enforces the divergence-free condition. We show that this case also yields a positive definite Hamiltonian (up to a boundary term); this is the only clear instance so far in which the AKE both admits a well-posed initial value problem and a positive definite Hamiltonian. We argue that in a certain class of asymptotically flat spacetimes---using the notion of asymptotic flatness introduced in \cite{DuaHil19}---it is reasonable to expect solutions of the AKE to approximate Killing vectors in the asymptotic region for an appropriate choice of initial data (though we do not yet have a full proof). Nevertheless, these results suggest that under the appropriate restrictions, the AKE may be useful for constructing semi-Killing vectors in asymptotically flat spacetimes.

This paper is organized as follows. After establishing conventions and general assumptions in Sec. \ref{sec:Conv}, we introduce in Sec. \ref{sec:AKEGaussLaw} the AKE, its Komar current, and discuss the conditions under which it admits a Gauss law. In Sec. \ref{sec:AKVnotAKE}, we show that, in general, the AKE admits solutions that do not approximate Killing vectors even in spacetimes that admit Killing vectors, and that an appropriate choice of initial data are needed. A Hamiltonian analysis is performed in Sec. \ref{sec:Stability}, in which we discuss the unboundedness of the Hamiltonian. We then address the hyperbolicity of the AKE in Sec. \ref{sec:Hyperbolicity}.  Afterwards, in Sec. \ref{sec:Vm2C}, we discuss how in spite of the preceding results, the AKE can simultaneously yield a well-posed initial value problem and positive-definite Hamiltonian in vacuum spacetimes. In Sec. \ref{sec:AKV}, we conclude with a general discussion concerning the suitability of the AKE for constructing semi-Killing vectors.


%
%

\section{Conventions} \label{sec:Conv}
Let $(\mathcal{M},g)$ denote a four-dimensional spacetime where $g$ is a Lorentzian metric of signature $(-,+,+,+)$. Throughout this article, greek indices will be used as spacetime coordinate indices. Let $n^{\mu}$ denote the unit normal ($n_\mu n^\mu=-1$) to a three-dimensional spacelike hypersurface $\Sigma\subset \mathcal{M}$. As is customary, lower case latin indices from the second half of the alphabet $\{i,j...\}$ will represent spatial coordinate indices.  We employ a 2+1 split on $\Sigma$; in this regard, we adhere to the conventions of \cite{Hil15}: $s^i$ with $s_is^i=1$ will denote the normal vector to a two-dimensional surface $\mathcal{S}\subset \Sigma$ and the corresponding induced metric will be denoted $q_{ij}$. Consistent with these conventions, one has the projection tensor
\begin{equation}\label{2p1p1splitmetric}
  g_{\mu\nu}= q_{\mu\nu}-n_{\mu}n_{\nu}+s_{\mu}s_{\nu}.
\end{equation}

\noindent Additionally, upper case latin indices $\{A,B...\}$ will be used as coordinate indices associated to a fiduciary coordinate system $\theta^A$ on $\mathcal{S}$---see \cite{Hil15} for further discussion on the 2+1+1 split. Our curvature conventions are fixed by
\begin{equation}\label{AKEGL-commutatorCD}
\nabla_{\mu}\nabla_{\nu}v^{\sigma} -  \nabla_{\nu}\nabla_{\mu}v^{\sigma} =  R^{\sigma}{}_{\lambda \mu\nu} v^{\lambda}.
\end{equation}

\noindent The divergence of a spacetime vector $V^\mu$ will be written in the following way:
\begin{equation}\label{AKEGL-DivergenceNotation}
\nabla \cdot V = \nabla_\mu V^\mu .
\end{equation}
\noindent The domain of dependence of a spacelike region $\mathcal{U}\subseteq \Sigma$ will be denoted by $\mathcal{D}(\mathcal{U})$.


%
%
\section{AKE and Gauss law} \label{sec:AKEGaussLaw}

The AKE \cite{Taubes1978,Bonaetal2005} takes the form,
\begin{equation} \label{AKEGL-AlmostKillingEquation}
\Box \xi^{\nu} + R{^\nu}{_\sigma} \, \xi^{\sigma} + (1 - \mu) \nabla^\nu \left(\nabla \cdot \xi \right) = 0 ,
\end{equation}

\noindent where $\mu$ is a constant parameter. Note that when $\mu=2$, the time derivatives of the $\xi^0$ component drop out entirely; as we shall see later, this is a parameter choice for which the AKE fails to be hyperbolic. The definition of almost Killing vectors via solutions to Eq. \eqref{AKEGL-AlmostKillingEquation} can be motivated by the observation that if $\zeta^\mu$ is a Killing vector then $\zeta^\mu$ satisfies the wave equation (though $\nabla \cdot \zeta=0$ for a Killing vector, we keep it here to illustrate the motivation for the last term in the AKE)
\begin{equation} \label{AKEGL-WaveEqMu1}
 \square \zeta^{\mu} + R^{\mu}{}_{\nu} \zeta^{\nu} + \nabla^\nu \left(\nabla \cdot \zeta \right)
  = 0.
\end{equation}
The latter follows simply by taking the divergence of the Killing equation (KE)
\begin{equation}\label{AKEGL-KE}
\nabla_\mu\zeta_\nu +\nabla_\nu\zeta_\mu = 0.
\end{equation}

\noindent Explicit solutions to the AKE have been constructed for some spacetimes---see the examples in \cite{Ruizetal2014}.  In \cite{FengCurrents2018}, a solution to the $\mu=2$ AKE was constructed for a Vaidya spacetime describing the emission of a spherical pulse of radiation from a star; this solution has the property that it approximately satisfies the Killing condition everywhere away from the pulse.

Observe that the associated Komar current of an almost-Killing vector is given by,
\begin{equation} \label{AKEGL-KomarAK}
	J^\nu = 2R^{\nu}{_\sigma} \xi^\sigma + (2-\mu) \nabla^\nu \left(\nabla \cdot \xi \right).
\end{equation}
\noindent Note that if the solutions to the AKE satisfy $\nabla \cdot \xi=0$ or if $\mu=2$, the Komar current vanishes in a vacuum.
The second term in Eq. \eqref{AKEGL-KomarAK} can in fact be rewritten in terms of the Lie derivative along $\xi$ of the Levi-Civita connection for $g_{\mu \nu}$---see \cite{ChaPad15}. This term would trivially vanish if $\xi$ were a Killing vector.

\definecolor{dgreen}{HTML}{008000}
\definecolor{fred}{HTML}{ff0000}
\definecolor{dred}{HTML}{d40000}
\definecolor{fbrown}{HTML}{6c6753}
\definecolor{pblue}{HTML}{2c5aa0}
\begin{figure}
\includegraphics[width=0.45\textwidth]{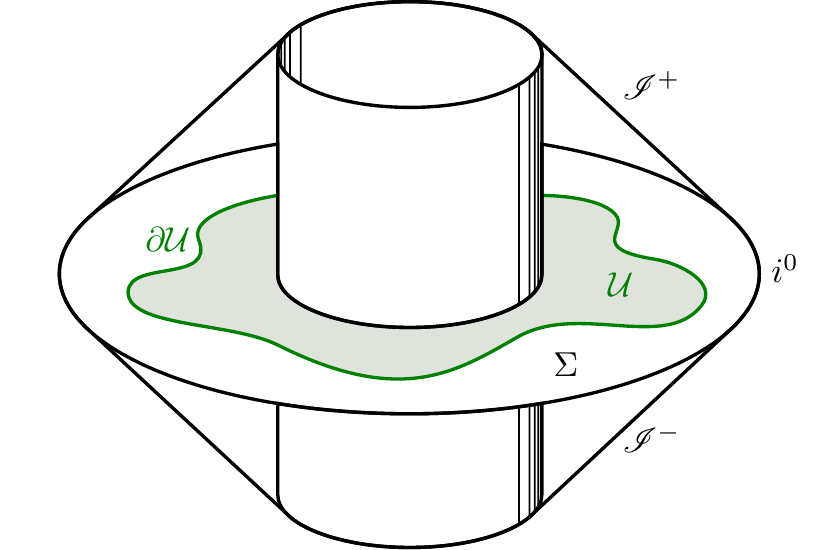}
\caption{An illustration of the vacuum region of an asymptotically flat spacetime. Matter (in particular, Ricci curvature) is assumed to have compact support on all spatial slices ($\Sigma$ is an arbitrary spatial slice), and the region containing matter and/or black holes has been excluded from the diagram. The shaded region is $\mathcal{U}\subset \Sigma$, with boundary $\partial \mathcal{U}$. Since the Komar current vanishes in a vacuum for the parameter choice $\mu=2$ or for solutions satisfying $\nabla \cdot \xi= 0$, the resulting surface integrals $Q_{\mathcal{U}}$ (Eq. (\ref{AKEGL-KomarSurfIntU})) will be independent of the choice of boundary $\partial \mathcal{U}$ for the region $\mathcal{U}$. Observe that one can deform the region $\mathcal{U}$ in a timelike direction, and that the value of $Q_{\mathcal{U}}$ must remain unchanged if $\partial \mathcal{U}$ is held fixed. It follows that if $\partial \mathcal{U}$ remains in the vacuum region, $Q_{\mathcal{U}}$ is independent of the local geometry of $\partial \mathcal{U}$.} \label{fig:AsympFlatMatter}
\end{figure}

Additionally, for nonvacuum solutions in which $R_{\mu \nu} \, \xi^\mu$ has compact support on a hypersurface $\Sigma$, one can formulate a Gauss law for matter from solutions to the AKE, provided that $\mu=2$ or the solutions to the AKE satisfy $\nabla \cdot \xi=0$. The argument is illustrated in Fig. \ref{fig:AsympFlatMatter}, and assumes that the boundary $\partial \mathcal{U}$ for the compact surface $\mathcal{U} \subset \Sigma$ lies in the vacuum region of an asymptotically flat spacetime. Since the Komar current vanishes in a vacuum, the integral:
\begin{equation} \label{AKEGL-KomarSurfIntU}
	Q_\mathcal{U} = 2 \int_{\partial \mathcal{U}} \nabla^{[\mu} \xi^{\nu]}\, dS_{\mu \nu}
\end{equation}

\noindent is independent of the local geometry of the boundary surface $\partial \mathcal{U}$, depending only on the amount of Ricci curvature (in particular the quantity $R_{\mu \nu} \, \xi^\mu \, n^\nu$) that the surface $\partial \mathcal{U}$ encloses. We observe that if the boundary surface $\partial \mathcal{U}$ lies in a vacuum region as described in Fig. \ref{fig:AsympFlatMatter}, the value of $Q_\mathcal{U}$ is independent of both the spatial and temporal placement of $\partial \mathcal{U}$, so long as there exists a deformation of $\partial \mathcal{U}$ such that the boundary surface remains in a vacuum. This property of $Q_\mathcal{U}$ suggests that solutions of the AKE satisfying $\nabla \cdot \xi=0$ or $\mu=2$ yield a Gauss law in asymptotically flat spacetimes which, for asymptotically timelike solutions $\xi^\mu$, provides a measure of matter content in those spacetimes; as we will argue later, it can in fact be used to obtain a measure of nongravitational radiation if the energy-momentum tensor is trace free.

Of course, the formulation of such a Gauss law depends on whether the AKE admits solutions satisfying $\nabla \cdot \xi = 0$ or for the parameter choice $\mu=2$. We will later show that the AKE is hyperbolic and preserves the divergence-free condition in vacuum for the parameter choice $\mu=1$. For the $\mu=2$ case, the AKE is ill-posed and must be supplemented with the divergence-free equation $\nabla \cdot \xi=0$; as we shall see later, the system consisting of the $\mu=2$ AKE and the equation $\nabla \cdot \xi=0$ can be reformulated as an initial value problem for the $\mu=1$ AKE with divergence-free initial data [in particular initial data satisfying $(\nabla \cdot \xi)|_\Sigma=0$ and $\mathcal{L}_n( \nabla \cdot \xi)|_\Sigma = 0$]. In an asymptotically flat spacetime, the Gauss law depends on whether one can smoothly extend such vacuum solutions to divergence-free solutions of the $\mu=1$ AKE over regions where $R_{\mu \nu} \neq 0$.

\begin{figure}
\includegraphics[width=0.45\textwidth]{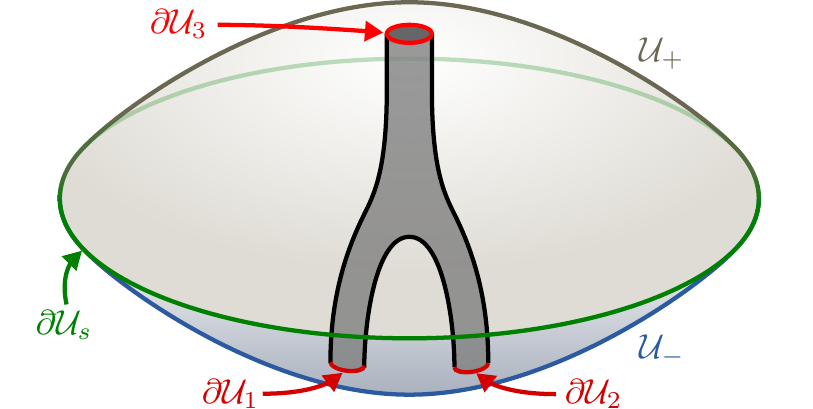}
\caption{The boundary of a vacuum spacetime region around a pair of merging black holes. The inner boundary surface is described by the well-known ``pair of pants'' diagram (\cite{HawkingEllis,MTW,Matzneretal1995}) for the horizons of merging black holes. The outer boundary surface consists of two spacelike surfaces $\mathcal{U}_{-}$ and $\mathcal{U}_{+}$ with respective boundaries
$\partial \mathcal{U}_{-} = \ipartial \mathcal{U}_{1} \cup \ipartial \mathcal{U}_{2} \cup \ipartial \mathcal{U}_{s}$
and
$\partial \mathcal{U}_{+} = \ipartial \mathcal{U}_{3} \cup \ipartial \mathcal{U}_{s}$. } \label{fig:PantsDiagram}
\end{figure}

Irrespective of the extendability of the divergence-free vacuum solutions of the AKE to nonvacuum regions, one can still formulate a Gauss law for a system of black holes in vacuum. In particular, divergence-free vacuum solutions of the AKE yields a dynamically conserved quantity that does not depend on the behavior of the black holes---it holds even for binary black hole mergers. In particular, one chooses $\partial \mathcal{U}$ to consist of black hole horizons\footnote{Here, we introduce the notation $\ipartial$ to denote a subset of a boundary surface.} $\ipartial \mathcal{U}_i$ and a spatial boundary $\ipartial \mathcal{U}_s$ which encloses the black holes. The surface integrals over the black hole horizons (which may be apparent horizons or dynamical/isolated horizons \cite{AshtekarKrishnan2004,*AshtekarKrishnan2003,*AshtekarKrishnan2002}) are then conserved, since one can choose $\ipartial U_s$ to be located at spatial infinity $i^0$, so that it forms the boundary for all spacelike hypersurfaces. As illustrated in Fig. \ref{fig:PantsDiagram}, the surface integral over $\ipartial \mathcal{U}_s$ does not depend on the local geometry of $\mathcal{U}$, and since $Q_\mathcal{U}=0$, the sum of the surface integrals over the black hole horizons $\ipartial \mathcal{U}_i$ must be constant; the Gauss law does not depend on the geometry or topology of black hole horizons.

The Gauss law discussed here comes with an important caveat: given a timelike solution to the AKE, $Q_{\mathcal{U}}$ cannot yield a satisfactory notion of gravitational energy, since it is insensitive to the presence of gravitational radiation. To see this, consider a binary neutron star system, which loses energy\footnote{\label{footnote:energymom}In the absence of a good local definition of energy and momentum for the vacuum gravitational field, we define these notions operationally for the sake of the present discussion. In particular, we define the energy and momentum for the gravitational field to be its capacity to do work and impart momentum, respectively.} by emitting gravitational radiation. If $\partial \mathcal{U}$ remains in a vacuum, the value of $Q_\mathcal{U}$ will not change under time evolution\footnote{In particular, under the flow generated by $\partial / \partial t$.} even as gravitational radiation exits the region $\mathcal{U}$. Thus, we conclude that $Q_\mathcal{U}$ cannot measure the energy contained in the gravitational field.

What then, does $Q_\mathcal{U}$ measure? In \cite{Wald94,*IyeWal94}, the integral (\ref{AKEGL-KomarSurfIntU}), evaluated on a stationary black hole horizon, was identified as black hole entropy for the future-directed horizon Killing field.\footnote{We refer the reader to \cite{Jacobsonetal1994} for further discussion of the difficulties with the original interpretation of entropy given in \cite{Wald94,*IyeWal94}.} However, one immediately encounters a difficulty with this interpretation for solutions of the AKE in dynamical vacuum spacetimes. The difficulty occurs for the same reasons discussed in the preceding paragraph; $Q_\mathcal{U}$ is insensitive to gravitational radiation, so $Q_\mathcal{U}$ cannot increase when absorbing gravitational radiation, as one might expect for the black hole entropy. However, it does satisfy an increase law when absorbing matter, so it may suffice as a partial measure of entropy that does not count states associated with gravitational degrees of freedom.

Aside from the entropy interpretation, the quantity $Q_\mathcal{U}$ does seem to provide a measure of matter content for asymptotically timelike solutions $\xi^\mu$ to the almost-Killing equation, provided that $\xi^\mu$ is divergence free or $\mu=2$. To see this, note that $Q_\mathcal{U}$ may be written as an integral of $R_{\mu \nu} \, \xi^\mu \, n^\nu$ over $\mathcal{U}$, and combined with the (trace-reversed) Einstein equations, $Q_\mathcal{U}$ takes the form:
\begin{equation} \label{AKEGL-KomarSurfIntMatterContent}
	Q_\mathcal{U}=\int_{\mathcal{U}} \left(T_{\mu \nu} - \frac{1}{2} T \, g_{\mu \nu}\right) \xi^\mu \, n^\nu d\Sigma ,
\end{equation}

\noindent where $T:=T_{\mu \nu} \, g^{\mu \nu}$. We observe that if $\xi^\mu \propto n^\mu$, then $Q_\mathcal{U}$ is positive definite if the strong energy condition is satisfied; however one should keep in mind that there exist classical matter configurations that violate the strong energy condition \cite{HawkingEllis,VisserBarcelo2000}. In this sense, we may regard $Q_\mathcal{U}$ to be a measure of matter content in spacetime.

This is indeed the case when $\xi$ is a timelike Killing vector in an asymptotically flat spacetime; $Q_{U}$ is then Komar mass (up to the infamous factor of 2). One might then be tempted to define a notion of gravitational energy for an asymptotically flat spacetime by taking the difference between the Arnowitt-Deser-Misner (ADM) mass and Eq. (\ref{AKEGL-KomarSurfIntMatterContent}). However, as we have remarked, there exist classical matter configurations that violate the strong energy condition, so $Q_\mathcal{U}$ is not necessarily positive definite. Furthermore, the usefulness of such a definition is limited, as the gravitational energy defined this way will satisfy an absolute conservation law, since the ADM mass and $Q_\mathcal{U}$ are conserved; one cannot describe energy transfer between matter and the gravitational field with such a definition. For the reasons listed here and in the preceding paragraphs, we find it more appropriate to interpret the quantity $Q_\mathcal{U}$ computed for asymptotically timelike $\xi^\mu$ to be some measure of matter content rather than a measure of total mass, total energy, or total entropy; this interpretation is in line with \cite{Lynden-BellBicak2017}, in which it was argued that if $\xi^\mu$ is a timelike Killing vector,  Eq. \eqref{AKEGL-KomarSurfIntMatterContent} provides a measure of effective gravitating mass. For a spacetime filled with nongravitational radiation ($T=0$), $Q_\mathcal{U}$ does have a proper interpretation as radiation content for timelike $\xi^\mu$, as $Q_\mathcal{U}$ is the integral of the energy as defined by $T_{\mu \nu} \xi^\mu n^\nu$. This was explored in \cite{FengCurrents2018}, in which $Q_\mathcal{U}$ was explicitly computed for the Vaidya spacetime solution discussed therein.


%
%
\section{Almost-Killing and Killing vectors} \label{sec:AKVnotAKE}
Solutions of the AKE are generalizations of Killing vectors since Killing vectors are themselves solutions to the AKE. However, one might ask whether solutions of the AKE can be regarded as approximately Killing in some sense. We consider this issue by rewriting the AKE, introducing the reduction variables,
\begin{equation}\label{AKEGL-Qdef}
Q_{\mu\nu} = \nabla_{(\mu}\xi_{\nu)}, \quad P_{\mu\nu} = \nabla_{[\mu}\xi_{\nu]}.
\end{equation}
It is straightforward to show that the divergence of $Q_{\mu\nu}$ satisfies
\begin{equation} \label{AKEGL-Qdiv}
\nabla^{\mu}Q_{\mu}{}^{\nu} =  \tfrac{1}{2} \square \xi^{\nu} + \tfrac{1}{2} R^{\nu}{}_{\mu} \xi^{\mu}  + \tfrac{1}{2} \nabla^{\nu}\nabla_{\mu}\xi^{\mu}.
\end{equation}
Defining $\omega:=Q_{a}{}^{a}$, one may use \eqref{AKEGL-Qdiv} to rewrite the AKE \eqref{AKEGL-AlmostKillingEquation} in the first order form
\begin{subequations}\label{AKEGL-firstorderAKE}
  \begin{align}
  & \nabla_{\mu}\xi_{\nu} - \nabla_{\nu}\xi_{\mu} - 2 P_{\mu\nu} =0, \label{AKEGL-firstorderAKEc} \\
  & \nabla_{\mu}\xi_{\nu} + \nabla_{\nu}\xi_{\mu} - 2 Q_{\mu\nu} =0, \label{AKEGL-firstorderAKEa} \\
  & \nabla^{\nu}Q^{\mu}{}_{\nu} - \tfrac{1}{2} \mu \nabla^{\nu}\omega  =0. \label{AKEGL-firstorderAKEb}
\end{align}
\end{subequations}
This form of the AKE illustrates the differences between solutions to the AKE and the KE.  The most immediate observation is that Eq. \eqref{AKEGL-firstorderAKE} reduces to  Eq. \eqref{AKEGL-KE} if $Q_{\mu\nu}=0$. Additionally, notice that  Eq. \eqref{AKEGL-firstorderAKEb} is trivially satisfied if $Q_{\mu\nu}=R_{\mu\nu}$ and $\mu=1$ by virtue of the contracted second Bianchi identity. Similarly, for general values of $\mu$, Transverse-Trace free (TT) tensors constitute particular solutions to \eqref{AKEGL-firstorderAKEb}.  These observations show that in general, a solution to the wave equation \eqref{AKEGL-WaveEqMu1} does not necessarily correspond to a solution of \eqref{AKEGL-KE}---even in spacetimes which admit Killing vectors. In other words, Eq. \eqref{AKEGL-WaveEqMu1} is a necessary but not sufficient condition for the existence of a Killing vector in the spacetime. This behavior is not unexpected as the sufficiency in the latter argument is obtained by prescribing appropriate initial data, known as Killing initial data (KID), for the wave equation \eqref{AKEGL-WaveEqMu1}---for a complete discussion on the KID equations see \cite{Mon76, Coll77, GarKha19}. Finally, notice that, in general, solutions to the AKE do not necessarily approximate Killing vectors. To see this, note that Eq. (\ref{AKEGL-firstorderAKEb}) is satisfied by any TT tensor $Q_{\mu \nu}$ the components of which need not be small. For this reason, the term ``almost Killing'' is somewhat of a misnomer, but since it appears in the existing literature, we continue its use in this article.

%
%
\section{Runaway instability} \label{sec:Stability}
Since solutions of the AKE do not approximate Killing vectors, one might ask whether the quantity $\nabla_{(\mu}\xi_{\nu)}$ remains small for solutions of the AKE, given some notion of approximate Killing initial data. This is a difficult question to answer fully, but we demonstrate in this section that one can exploit the Hamiltonian structure of the AKE to identify and exclude situations in which the solutions of the AKE are potentially subject to instabilities and runaway behavior.\footnote{The nature of the instability we describe here is similar to that of the Ostrogradsky instability \cite{Ostrogradsky1850,Woodard2015}, but it is not strictly an Ostrogradsky instability because the action we use (\ref{AKEGL-Action}) does not contain higher derivatives.} A Hamiltonian analysis is possible because the AKE can be derived from a variational principle. The AKE follows from the action \cite{Bonaetal2005},
\begin{equation}\label{AKEGL-Action}
S[\xi]=-\frac{1}{2}\int_U \left( 2 \nabla^{(\mu} \xi^{\nu)} \, \nabla_{(\mu} \xi_{\nu)} - \mu ( \nabla \cdot \xi )^2 \right) dV .
\end{equation}

\noindent Note that the action and Lagrangian density vanishes when $\xi^\mu$ is a Killing vector. To obtain an expression for the Hamiltonian, we employ Weiss variation methods, a more complete discussion of which can be found in the Refs. \cite{FengMatznerWeiss2018,SudarshanCM,MatznerShepleyCM}. The Weiss variation is obtained by performing displacements of the boundary, and allowing for field variations at the boundary surfaces. Including boundary terms and the variation with respect to the metric, the variation of the action takes the following form ($\mathbb{E}^\nu[\cdot]$ being the Euler-Lagrange operator):
\begin{equation}\label{AKEGL-ActionVarFullA}
\begin{aligned}
\delta S =& \int_U \mathbb{E}^\nu[\xi] \delta \xi_\nu\delta g^{\mu \nu} \, dV  +  \varepsilon  \int_{\partial U} \mathcal{L} \, \delta x^\nu \, n_\nu \, d\Sigma\\
& -  \int_{\partial U} \left[\left(2 \, \nabla^{(\mu} \xi^{\nu)} - \mu \, g^{\mu \nu}\, \nabla \cdot \xi \right) \delta \xi_{\mu} \right] \varepsilon \, n_\nu \, d\Sigma ,
\end{aligned}
\end{equation}

\noindent where $\delta x^\nu$ is the boundary displacement, and $\mathcal{L}$ is a Lagrangian density of the form,
\begin{equation}\label{AKEGL-LagDens}
\mathcal{L}=-\frac{1}{2} \left( 2 \nabla^{(\mu} \xi^{\nu)} \, \nabla_{(\mu} \xi_{\nu)} - \mu ( \nabla \cdot \xi )^2 \right).
\end{equation}

\noindent We identify the (covariant) polymomentum,
\begin{equation}\label{AKEGL-Polymomentum}
\begin{aligned}
P^{\mu \nu} := \frac{\partial \mathcal{L}}{\partial (\nabla_\mu \xi_\nu)} = -  2 \, \nabla^{(\mu} \xi^{\nu)}  + \mu \, g^{\mu \nu}\, \nabla \cdot \xi
\end{aligned}
\end{equation}

\noindent which may be used to simplify the variation of the action,
\begin{equation}\label{AKEGL-ActionVarFullA2}
\begin{aligned}
\delta S =& \int_U \, \mathbb{E}^\nu[\xi] \delta \xi_\nu \, dV +  \int_{\partial U} \left[ P^{\mu \nu} \, \delta \xi_{\mu} + \mathcal{L} \, \delta x^\nu\right] \varepsilon \, n_\nu \, d\Sigma.
\end{aligned}
\end{equation}

We now wish to obtain the Weiss form of the variation. Following the approach outlined in \cite{FengMatznerWeiss2018}, we define the total variation of $\xi^\nu$ on the displaced boundary surface $\partial U^\prime$,
\begin{equation}\label{AKEGL-TotalVariationxi}
\begin{aligned}
\Delta \xi_\mu :&= (\xi_\mu + \delta \xi_\mu)|_{\partial U^\prime} - \xi_\mu |_{\partial U} \\
&= (\delta \xi_\mu + \pounds_{\delta x} \xi_\mu)|_{\partial U}.
\end{aligned}
\end{equation}

\noindent The variation of the action takes the form,
\begin{equation}\label{AKEGL-ActionVarFullA3}
\begin{aligned}
\delta S =& \int_U \, \mathbb{E}^\nu[\xi] \delta \xi_\nu \, dV \\
& +  \int_{\partial U} \left[ P^{\mu \nu} \, \Delta \xi_{\mu} - P^{\mu \nu} \, \pounds_{\delta x} \xi_\mu+ \mathcal{L} \, \delta x^\nu\right] \varepsilon \, n_\nu \, d\Sigma.
\end{aligned}
\end{equation}

\noindent Now we consider boundary surfaces defined by constant values of coordinate time $t$. If the boundary displacement vector is proportional to the coordinate basis vector $\partial / \partial t$, so that $\delta x^\mu = \Delta t \, ({\partial}/{\partial t})^\mu$ (where $\Delta t$ is a constant coordinate time displacement of the boundary), the Lie derivative operator becomes $\pounds_{\Delta t \, \partial / \partial t} = \Delta t \, \partial / \partial t$, so that:
\begin{equation}\label{AKEGL-ActionVarFullA4}
\begin{aligned}
\delta S =& \int_U \, \mathbb{E}^\nu[\xi] \delta \xi_\nu \, dV + \varepsilon \, \int_{\partial U} P^{\mu \nu} \, \Delta \xi_{\mu} \, n_\nu \, d\Sigma\\
& - \Delta t  \int_{\partial U} \left[\varepsilon \, P^{\mu \nu} \, n_\nu \, \dot{\xi}_\mu - \alpha \, \mathcal{L} \right] \, d\Sigma,
\end{aligned}
\end{equation}

\noindent where $\alpha := \varepsilon \, n_\mu \, (\partial /\partial t)^\mu$ is the ADM lapse function. Given the Weiss variation (\ref{AKEGL-ActionVarFullA4}), we may identify the conjugate momentum field $\pi^\mu$ and the Hamiltonian $H$,
\begin{equation}\label{AKEGL-ConjMomentum}
\begin{aligned}
\pi^\mu = \varepsilon \, P^{\mu \nu} \, n_\nu
\end{aligned}
\end{equation}
\begin{equation}\label{AKEGL-AKEHamiltonian}
\begin{aligned}
H[\pi^\cdot,\xi_\cdot,\gamma_{\cdot \cdot},\alpha,\beta^\cdot] = \int_{\Sigma}\mathcal{H} \, d\Sigma.
\end{aligned}
\end{equation}

\noindent where $\alpha$ and $\beta^i$ are the ADM lapse and shift \cite{MTW}, $\gamma_{ij}$ is the 3 metric on the hypersurface $\Sigma$, and $\mathcal{H}$ is the Hamiltonian density given by
\begin{equation}\label{AKEGL-AKEHamiltonianDensity}
\begin{aligned}
\mathcal{H} = \pi^\mu \, \dot{\xi}_\mu - \alpha \, \mathcal{L} .
\end{aligned}
\end{equation}

\noindent Note that the Hamiltonian $H$ vanishes when $\xi^\mu$ becomes Killing. To see this, note that when $\xi^\mu$ is a Killing vector, both the expression for the Lagrangian density $\mathcal{L}$ in Eq. (\ref{AKEGL-LagDens}) and the expression for $P^{\mu \nu}$ in Eq. (\ref{AKEGL-Polymomentum}) vanishes.

That the Hamiltonian is unbounded below can be seen by computing it in the orthonormal basis,  assuming Gaussian normal coordinates and with timelike direction aligned with $\partial / \partial t=n$. In this basis, the Hamiltonian takes the form,
\begin{equation}\label{AKEVar-HamDensOB}
\begin{aligned}
\mathcal{H}_o
	=& \frac{1}{2} \biggl((\mu-2) \dot{\xi}_{0} \, \dot{\xi}_{0} + \dot{\xi}_{j} \, \dot{\xi}^{j} - \nabla_{i} \xi_{0} \, \nabla^{i} \xi_{0} + 2 \nabla_{(i} \xi_{j)} \, \nabla^{(i} \xi^{j)}\\
&\>\>\>\>\>\>\>\> - \mu \, \nabla_{i} \xi^{i} \, \nabla_{j} \xi^{j}  \biggr) .
\end{aligned}
\end{equation}

\noindent Here, we see that when $\mu < 2$, one of the kinetic terms has the wrong sign, which correspond to ghost modes in the system. When $\mu > 2$, there are no ghosts, but there are two terms which are generally negative; one can show by way of a trace decomposition of $\nabla_{(i} \xi_{j)}$ that in general, the last two terms can be negative if $\mu>1/3$. The unboundedness of the Hamiltonian is potentially disastrous, since one can in general have runaway solutions which can produce uncontrolled growth in $\nabla_{(\mu} \xi_{\nu)}$, driving the solutions far from the Killing condition.

Of course, these arguments do not constitute a rigorous proof that the solutions will exhibit runaway behavior---we will in fact discuss an exception in the next section---but they bring to light an issue that should be addressed when discussing the long-term behavior of solutions to the AKE. Before discussing the exceptional $\mu=2$ case, we briefly examine several alternate scenarios in which a system with an unbounded Hamiltonian can avoid runaway solutions. In the ghost-free case $\mu > 2$, one possibility is that there exist situations where the potential terms in the Hamiltonian have local minima, so that the system is metastable,\footnote{See \cite{Salvio2019} for examples of metastability in systems with unbounded Hamiltonians.} and in this way, one can avoid runaway behavior for the appropriate initial data and spacetime geometries. If the system has ghosts (the $\mu < 2$ case), there are two possibilities. It has been suggested \cite{Sbisa2014} that in a system with ghosts, derivative interactions tend to stabilize the system; we note that the interactions in the AKE Lagrangian consist of derivative interactions. The second possibility is that the system becomes stable against such runaway behavior if one can somehow decouple the ghost modes (see, for instance, \cite{Salvio2019,PaisUhlenbeck1950,*Pavsic2016,*Salvio2016}). The expression for $\mathcal{H}_o$  in Eq. (\ref{AKEVar-HamDensOB}) does seem to suggest that there may exist such a decoupling, and the negative term that does not involve $\xi_0$ can be dealt with by requiring that $1/3 < \mu < 2$. However, we stress that Eq. (\ref{AKEVar-HamDensOB}) is only valid in the orthonormal basis, and that in general, one has couplings between $\xi_0$ and $\xi_i$; we are not suggesting that a decoupling occurs between the components $\xi_0$ and $\xi_i$, but perhaps there may exist a change of variables or a coordinate gauge in which ghost variables decouple (we leave this matter for future work).


%
%
\section{Hyperbolicity}\label{sec:Hyperbolicity}
In this section, we investigate the hyperbolicity of the AKE.  In   particular, we show that the AKE is weakly hyperbolic for $\mu\neq 2$ and that $\mu=1$ is the only parameter choice for which the AKE is strongly hyperbolic.

We employ the methods of hyperbolicity analysis for second order systems, particularly those of \cite{Gun06}---see also \cite{Hil13,GKO2013,BonaPal2004,*BonaPal2005}. Using the symbol $\simeq$ to denote equality up to terms not included in the principal part, Eq. \eqref{AKEGL-AlmostKillingEquation} may be rewritten as
  \begin{equation}
  g^{\alpha\beta}\partial_\alpha\partial_\beta \xi^\mu + (1-\mu)g^{\mu\alpha}\partial_\alpha\partial_\beta\xi^\beta  \simeq 0 .
  \end{equation}
\noindent Using Eq. \eqref{2p1p1splitmetric}  and defining $\xi^n := n_\mu\xi^\mu$, $\xi^s := s_\mu \xi^\mu$ and $\xi^A := q_{\mu}{}^{A}\xi^{\mu}$, we obtain
  \begin{flalign}
     & (\mu-2)\partial_n^2\xi^n + \partial_s^2\xi^n + (1-\mu)\partial_n\partial_s \xi^s \nonumber\\  & \qquad \qquad  + (1-\mu)\partial_n\partial_A \xi^A
     + q^{AB}\partial_A\partial_B \xi^n  \simeq 0,\\
    & \partial_n^2 \xi^s -(2-\mu)\partial_s^2\xi^s + (1-\mu)\partial_s\partial_n\xi \nonumber\\ &  \qquad \qquad  -(1-\mu)\partial_s\partial_A\xi^A   - q^{AB}\partial_A\partial_B\xi^s \simeq 0, \\
    & \partial_n^2\xi^A -\partial_s^2 \xi^A - q^{CD}\partial_C\partial_D \xi^A  \simeq 0,
  \end{flalign}
\noindent where we use the shorthand notation $n^\mu\partial_\mu=\partial_n$ and $s^\mu\partial_\mu=\partial_s$. From the latter expression, one can read off the principal symbol $P^s$ of the system as defined in \cite{Gun06} for second order systems---see Sec. 1.5 in \cite{Hil13} for a concise discussion. The principal symbol $P^s$ takes the form,
  \begin{equation}
    P^s =
  \begin{pmatrix}
O & I \\
A & B
  \end{pmatrix} ,
  \end{equation}

\noindent where $O$ and $I$ denote the $4\times4$ zero and identity matrices, respectively, and
   \begin{equation}
    A =
   \begin{pmatrix}
   \frac{1}{2-\mu} & 0 & 0 & 0 \\
   0 & 2-\mu & 0 & 0 \\
   0 & 0 & 1 & 0 \\
   0 & 0 & 0 & 1
   \end{pmatrix} \quad
   B=
   \begin{pmatrix}
   0 & \frac{\mu-1}{\mu-2} & 0 & 0 \\
   \mu-1 & 0 & 0 & 0 \\
   0 & 0 & 0 & 0 \\
   0 & 0 & 0 & 0
   \end{pmatrix} .
   \end{equation}

\noindent A direct calculation shows that $P^s$ has real eigenvalues if $\mu\neq 2$, but only possesses a complete set of eigenvectors if $\mu=1$. Consequently, the AKE is only strongly hyperbolic only for $\mu=1$ and merely weakly hyperbolic for any other choice $\mu \neq 2$.


%
%
\section{Divergence-free case} \label{sec:Vm2C}
The AKE is only strongly hyperbolic for $\mu=1$, but from Sec. \ref{sec:Stability}, we found that this parameter choice corresponds to ghost degrees of freedom in the AKE. At this point, it seems that the AKE cannot be both strongly hyperbolic and ghost free. However, it turns out that there is one instance in which we can still ``have our cake and eat it too.'' In particular, we describe here a case in which the AKE yields a well-posed initial value problem and admits a Hamiltonian that is positive definite up to boundary terms.

As noted in \cite{Bonaetal2005}, the AKE becomes formally identical to the homogeneous Maxwell equations in a vacuum spacetime for the parameter choice $\mu=2$,
\begin{equation} \label{AKEGL-AlmostKillingEquationVacMax}
\Box \xi^{\nu} - \nabla^\nu \left(\nabla \cdot \xi \right) = 0 .
\end{equation}

\noindent The results of Sec. \ref{sec:Hyperbolicity} have established that the AKE is not hyperbolic for this parameter choice. In fact, it is straightforward to see why the Cauchy problem for the above equation is not well posed; one can easily verify that for any scalar field $\phi$, $\nabla_\mu\phi$ satisfies the above equation. Consequently, one can construct different solutions arising from the same initial data on $\Sigma$ provided that $\nabla_\mu\phi|_\Sigma=0$ and $\mathcal{L}_n\nabla_\mu\phi|_\Sigma=0$. One might expect that the $\mu = 2$ AKE generally suffers from the same problem when $R_{\mu \nu} \neq 0$. Fortunately, as noted in \cite{Bonaetal2005}, one can obtain a well-posed system by supplementing Eq. (\ref{AKEGL-AlmostKillingEquationVacMax}) with the equation,
\begin{equation} \label{AKEGL-AlmostKillingEquationConstraint}
\nabla \cdot \xi = 0 .
\end{equation}

\noindent Equations (\ref{AKEGL-AlmostKillingEquationVacMax}) and (\ref{AKEGL-AlmostKillingEquationConstraint}) constitute a system of equations which admit a well-posed initial value problem---this system is in fact mathematically equivalent to Maxwell's equations (in vacuum) for the vector potential in the Lorenz gauge. To see explicitly that (\ref{AKEGL-AlmostKillingEquationVacMax}) and (\ref{AKEGL-AlmostKillingEquationConstraint}) admit a well-posed initial value problem, observe that for $\mu = 1$ (in which the AKE is strongly hyperbolic), the identity $\nabla \cdot J=0$ for $J^\mu$ given by \eqref{AKEGL-KomarAK} may be used to obtain the propagation equation $\Box (\nabla \cdot \xi) = 0$ in a vacuum spacetime. If the initial data for the $\mu = 1$ AKE are such that\footnote{This condition on the initial data is typical of that used for Maxwell's equations in the Lorenz gauge---see for instance Chap. 10.2 of \cite{Wald}} $(\nabla \cdot \xi)|_\Sigma = 0$ and $\mathcal{L}_n (\nabla \cdot \xi)|_\Sigma = 0$, then the propagation equation ensures that the resulting solutions will satisfy $\nabla \cdot \xi = 0$. It follows that in a vacuum spacetime, Eqs. (\ref{AKEGL-AlmostKillingEquationVacMax}) and (\ref{AKEGL-AlmostKillingEquationConstraint}) may be reformulated as the $\mu = 1$ AKE with the initial data $(\nabla \cdot \xi)|_\Sigma = 0$ and $\mathcal{L}_n (\nabla \cdot \xi)|_\Sigma = 0$. One might observe that Eq. \eqref{AKEGL-AlmostKillingEquationConstraint} appears to remove $\mu$ from the AKE. However, requirement of strong hyperbolicity for the AKE demands $\mu=1$; Eq. \eqref{AKEGL-AlmostKillingEquationConstraint} becomes a dynamical constraint that is imposed at the level of initial data and is preserved by the evolution of the  $\mu=1$ AKE.

The point we wish to make here, however, is not just that Eqs. (\ref{AKEGL-AlmostKillingEquationVacMax}) and (\ref{AKEGL-AlmostKillingEquationConstraint}) yield a well-posed initial value problem, but rather to highlight that so far, Eqs. (\ref{AKEGL-AlmostKillingEquationVacMax}) and (\ref{AKEGL-AlmostKillingEquationConstraint}) form the only clear instance of the AKE which both yields a well-posed initial value problem and admits a positive definite Hamiltonian. The latter follows from the fact that in a vacuum spacetime, Eq. (\ref{AKEGL-AlmostKillingEquationVacMax}) can be derived from the Maxwell action, so that the action in Eq. (\ref{AKEGL-Action}) differs from the Maxwell action by a boundary term and an overall factor of 2. To see this, consider the Maxwell action, $S_M$,
\begin{equation}\label{AKEGL-ActionMaxwell}
\begin{aligned}
S_M
= \,&- \frac{1}{4} \int_U \, F^{\mu \nu} \, F_{\mu \nu} \, dV.
\end{aligned}
\end{equation}

\noindent Now assume a vacuum ($R_{\mu \nu}=0$) and perform an integration by parts, making use of the identity $\nabla^{\mu} \nabla_{\nu} V_{\mu} = R_{\mu \nu} V^{\mu} + \nabla_{\nu} (\nabla \cdot V)$, to rewrite Eq. (\ref{AKEGL-ActionMaxwell}) as
\begin{equation}\label{AKEGL-ActionMaxwellRW}
\begin{aligned}
S_M
	=\,& \frac{1}{4} \int_U \, \left(A^{\nu}  \, \Box A_{\nu} - A^{\nu} \, \nabla_{\nu} (\nabla \cdot A)\right) dV \\
	&- \int_{\partial U} \, A^{\nu} \, \nabla_{[\mu} A_{\nu]} \, d\Sigma^\mu ,
\end{aligned}
\end{equation}

\noindent An integration by parts for the AKE action $S$ yields the following expression:
\begin{equation}\label{AKEVar-ActionComp}
\begin{aligned}
S
	=\,& \frac{1}{2} \int_U \left(\xi^\nu \Box \xi_{\nu} + (1 - \mu)  \xi^\nu \nabla_\nu (\nabla \cdot \xi) \right) dV\\
	& - \frac{1}{2} \int_{\partial U} \left(2 \, \xi^{\nu} \, \nabla_{(\mu} \xi_{\nu)} - \mu \, \xi_\mu \nabla \cdot \xi \right) d\Sigma^\mu.
\end{aligned}
\end{equation}

\noindent Observe that upon setting $\mu=2$, the action $S$ is equivalent to the action $2 \, S_M$ up to boundary terms in a vacuum spacetime, and it follows that $S$ admits a positive definite Hamiltonian. We note that on shell, the Hamiltonian for the $\mu = 1$ case is also positive definite in the same sense. Since the $\mu = 1$ case with initial data $(\nabla \cdot \xi)|_\Sigma = 0$ and $\mathcal{L}_n (\nabla \cdot \xi)|_\Sigma = 0$ has been shown to be equivalent to the system in Eqs. (\ref{AKEGL-AlmostKillingEquationVacMax}) and
(\ref{AKEGL-AlmostKillingEquationConstraint}), we find that in a vacuum, the initial data $(\nabla \cdot \xi)|_\Sigma = 0$ and $\mathcal{L}_n (\nabla \cdot \xi)|_\Sigma = 0$ avoid the potential runaway instabilities of the generic $\mu = 1$ case.

One might worry that the boundary terms in the action \eqref{AKEVar-ActionComp} require fixing the derivatives of $\xi^\mu$ on the boundary in addition to $\xi^\mu$ itself, which would lead to an overdetermined variational principle. On the other hand, the procedure we have described here is often employed to relate the first and second order forms of kinetic terms in the action for relativistic field theories (see, for instance, Chap. 3 of \cite{Schwartz:2013pla}), so this procedure should still be valid for recovering local dynamics. These issues can be addressed in a simple manner: we point out that while fixing $\xi^\mu$ and its normal derivative on ${\partial U}$ is sufficient to guarantee the vanishing of boundary terms in the variation, this is not a necessary condition. In the general case, the requirement that the boundary term in the variation $\delta S$ vanish yields a constraint between the variations and their normal derivatives. More generally, different boundary terms in the action correspond to different boundary conditions in the variational principle, for instance, Neumann or Robin boundary conditions as opposed to the usual Dirichlet boundary conditions; note that boundary terms compatible with Neumann and Robin boundary conditions have been recently identified for Einstein gravity \cite{CKrishnan2016,*CKrishnan2017}.

Finally, since Eqs. (\ref{AKEGL-AlmostKillingEquationVacMax}) and (\ref{AKEGL-AlmostKillingEquationConstraint}) admit a well-posed initial value problem in a vacuum and, by construction $\nabla \cdot \xi=0$ and $\mu=2$, then the resulting Komar current yields an exact Gauss law as described earlier in this article. This result completes our claim that such solutions of the AKE can yield a Gauss law for vacuum black hole spacetimes, and that one can construct a dynamical invariant from the sum of surface integrals over black hole horizons constructed from such solutions of the AKE. It also provides further motivation for the investigation of the Gauss law in asymptotically flat spacetimes with nonvanishing Ricci curvature of compact support on spatial hypersurfaces.


%
%

\section{Can solutions of the AKE approximate Killing vectors?} \label{sec:AKV}

A natural question that arises is whether solutions to the AKE can
yield approximate Killing vectors, or at the very least, the semi-Killing vectors described in \cite{Komar1962} in a generic asymptotically flat spacetime. There certainly exist cases in which it does, for instance the Vaidya example in \cite{FengCurrents2018}, but here, we discuss whether one should expect (for an appropriate set of initial data) the AKE to yield some notion of approximate or semi-Killing vector in a general class of asymptotically flat spacetimes. We begin by revisiting the classical KID argument---see \cite{Coll77, Mon76}. Let $Q_{\mu\nu}$ be defined as in Eq. \eqref{AKEGL-Qdef}. If $\xi^\mu$ satisfies the wave equation $\Box \xi^\mu=0$, it can be shown that in a vacuum spacetime, $Q_{\mu\nu}$ satisfies the following propagation identity
  \begin{equation}\label{AKEGL-WaveEquationClassicalKIDS}
  \Box Q_{\mu\nu} = 2 R^\sigma{}_{\mu\nu}{}^{\lambda}Q_{\sigma\lambda} .
  \end{equation}

\noindent Then, if on a spacelike hypersurface $\mathcal{W}$ one imposes
\begin{equation}\label{AKEGL-preKID}
Q_{\mu\nu}|_{\mathcal{W}} = 0, \qquad \mathcal{L}_n Q_{\mu\nu}|_{\mathcal{W}} =0,
\end{equation}

\noindent a standard existence and uniqueness result for homogeneous wave equations ensures that $Q_{\mu\nu} =0$ in $\mathcal{D}(\mathcal{W})$.

\begin{figure}
\includegraphics[width=0.4\textwidth]{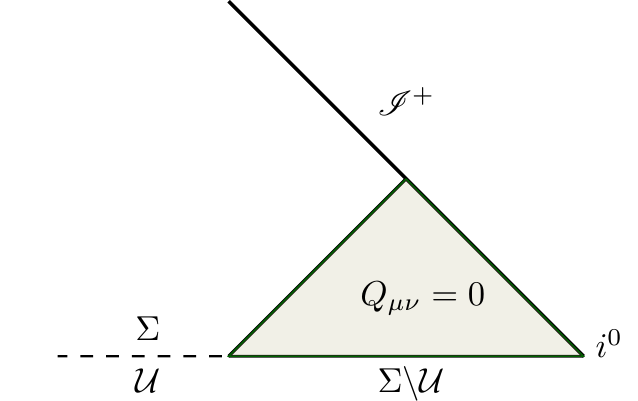}
\caption{Consider initial data for the wave equation \eqref{AKEGL-WaveEquationClassicalKIDS} in which $Q_{\mu \nu}$ and $\mathcal{L}_n Q_{\mu \nu}$ have compact support in $\mathcal{U} \subset \Sigma$, namely, assume that the KID equations $Q_{\mu \nu}=0$ and $\mathcal{L}_n Q_{\mu \nu}=0$ are satisfied on $\Sigma \textbackslash \mathcal{U}$. Then, Eq. \eqref{AKEGL-WaveEquationClassicalKIDS} ensures that the spacetime will admit a Killing vector in region represented by the shaded area in the diagram.
\label{fig:CompactSupport}}
\end{figure}

The latter means that if initial data for $\xi^{\mu}$ and $\mathcal{L}_n\xi^{\mu}$ are given such that Eqs. \eqref{AKEGL-preKID} are satisfied, then the solution to the propagation equation $\Box \xi^\mu=0$ will be a Killing vector in $\mathcal{D}(\mathcal{W})$---see \cite{Coll77, Mon76, GarKha19} for a detailed discussion. The latter equations are known as the Killing initial data (KID) equations, and they constitute necessary and sufficient conditions for the existence of Killing vectors. In Fig. \ref{fig:CompactSupport}, we illustrate the situation where the KID equations are only satisfied in the asymptotic region of an asymptotically flat and vacuum spacetime.  Here, it is assumed that initial data with compact support in $\mathcal{U} \subset \Sigma$ (where $\Sigma$ is a Cauchy hypersurface) for the wave equation \eqref{AKEGL-WaveEquationClassicalKIDS} are given. In other words, the conditions \eqref{AKEGL-preKID} are satisfied on $\mathcal{W}=\Sigma \textbackslash \mathcal{U}$. Such initial data can be constructed by gluing the data corresponding to the strong field region of a spacetime to the asymptotic end of initial data for a Kerr or a stationary spacetime---see \cite{Cor00,CorSch06}.  The classic KID argument ensures that the spacetime in the shaded region of the diagram admits a Killing vector.  Observe that if the propagation equation $\Box \xi^\mu=0$ is imposed, then the evolution of $Q_{\mu \nu}$ is governed by Eq. \eqref{AKEGL-WaveEquationClassicalKIDS} regardless of the initial data on $\mathcal{U}$. Consequently, if the initial data on $\mathcal{U}$ is small, i.e., slightly deviates from data satisfying the KID equations, then one could envision exploiting Eq. \eqref{AKEGL-WaveEquationClassicalKIDS} to show that $Q_{\mu \nu}$ remains small during evolution. Moreover, employing the definition of asymptotic flatness discussed in \cite{DuaHil19}, there exists a coordinate system $X^\mu$ close to $\mathscr{I}$ where  $g_{\mu\nu}=\eta_{\mu\nu}+O_p(R^{-1})$. Here $\eta_{\mu\nu}$ is the Minkowski metric, $p \geq 1$, and $O_p(R^{-m})$ means that partial derivatives $\partial_\mu$ of order $n$ decay as $O(R^{-m-n})$ for $0\leq n \leq p$. Using this definition with $p=2$, one concludes that $\Gamma^\mu{}_{\alpha \beta} = O(R^{-2})$ and $R_{\mu\nu\alpha\beta}=O(R^{-3})$.  Consequently, assuming $Q_{\mu\nu}$ is $O(1)$ and expanding Eq. \eqref{AKEGL-WaveEquationClassicalKIDS} in terms of partial derivatives, we get $\eta^{\alpha\beta}\partial_{\alpha}\partial_{\beta} Q_{\mu \nu} = O(R^{-1})$ close to $\mathscr{I}$. Neglecting the error terms and using the known falloff for solutions to the wave equation in flat space, one concludes that in fact $ Q_{\mu\nu} = O(R^{-1})$ close to $\mathscr{I}$; one may therefore expect the solutions of the AKE (for the appropriate initial data) to approximate Killing vectors in the neighborhood of spatial infinity of asymptotically flat spacetimes. Of course, we stress that these statements are not rigorous (they do not constitute a full proof), but they provide a reason to expect that the solutions of the AKE can yield semi-Killing vectors in a particular class of asymptotically flat spacetimes.


%
%
\section{Summary and future work}
In this article, we have shown that solutions of the almost-Killing equation admit a Gauss law for matter and black holes when $\nabla \cdot \xi =0$ or $\mu=2$. While the Gauss law is insensitive to gravitational radiation and cannot yield a definition for gravitational energy, we have argued that it may still useful for measuring matter content in asymptotically flat spacetimes.  We have studied the hyperbolicity of the AKE and have identified the parameter choice $\mu=1$ to be the only one for which the AKE is strongly hyperbolic, and we have shown that the remaining parameter choices $\mu\neq 2$ (which includes the $\mu=1/2$ for the ``conformal AKE'' in \cite{Taubes1978}) are only weakly hyperbolic. We have also performed a cursory Hamiltonian analysis for the AKE; in particular, we have found that the Hamiltonian is generally unbounded when $\mu \neq 2$ and has ghosts for $\mu < 2$. Fortunately, the Hamiltonian is positive definite up to boundary terms when $\mu=2$ in a vacuum, and when the AKE is supplemented with the constraint equation $\nabla \cdot \xi =0$ [which is equivalent to the $\mu=1$ case with divergence-free initial data $(\nabla \cdot \xi)|_\Sigma = 0$ and $\mathcal{L}_n (\nabla \cdot\xi)|_\Sigma = 0$], the system becomes well posed.

Finally, we argued that in asymptotically flat spacetimes, it is reasonable to expect that solutions of the AKE can yield semi-Killing vector solutions given appropriate restrictions in a neighborhood of $i^0$ for asymptotically flat spacetimes, though we are presently unable to fully prove this in full generality. The analysis given in this article therefore represents a first step towards a concrete realization of Komar's original notion of semi-Killing vectors in asymptotically flat spacetimes.


\begin{acknowledgments}
We thank Vitor Cardoso, Sumanta Chakraborty, David Hilditch, Masato Minamitsuji, and Simone Speziale for their helpful remarks and comments on this work. The authors would like to acknowledge networking support by the COST Action GWverse CA16104. JW thanks the hospitallity of CENTRA-IST during an academic visit where this work started. EG gratefully acknowledges support from IUCAA to fund an academic visit where part of this work was done, and also acknowledges support from the FCT (Portugal) IF Program No. IF/00577/2015. JCF acknowledges support from the FCT Grant No. PTDC/MAT-APL/30043/2017.
  \end{acknowledgments}


\bibliography{bibAKE}

\end{document}